\documentclass[aps,prx,longbibliography,superscriptaddress,nofootinbib,twocolumn]{revtex4-2}

\usepackage{dcolumn}

\usepackage{newtxtext}  
\usepackage{bm}         

\usepackage[english]{babel}

\usepackage[pdftex]{graphicx} 
\usepackage{float}
\usepackage{epstopdf}
\usepackage{booktabs}

\usepackage[dvipsnames,svgnames]{xcolor}

\usepackage[colorlinks=true,citecolor=blue,linkcolor=LimeGreen]{hyperref}

\usepackage{latexsym}
\usepackage{amsmath,amssymb,amsfonts,amsbsy,mathtools}
\usepackage{upgreek}    
\usepackage{mathrsfs}   
\usepackage{cancel}     
\usepackage{ulem}       
\usepackage{mathdots}   
\usepackage{setspace}   
\usepackage{dsfont}     
\usepackage{enumerate}  

\usepackage{tikz}
\usetikzlibrary{matrix}

\usepackage{appendix}

\usepackage{natbib}

\DeclarePairedDelimiter\bra{\langle}{\rvert}
\DeclarePairedDelimiter\ket{\lvert}{\rangle}
\DeclarePairedDelimiterX\braket[2]{\langle}{\rangle}{#1 \delimsize\vert #2}

\newcommand\kk{\mathbf{k}}
\newcommand\kkk{{\bm{\mathit{k}}}}
\newcommand\rr{\mathbf{r}}
\newcommand\RR{\mathbf{R}}

\newcommand\dip{\wp}                

\begin{document}
\title{
Long-range interactions in Weyl dense atomic arrays protected from dissipation and disorder
} 

\author{I\~naki Garc\'{i}a-Elcano}
\email{innaki.garciae@gmail.com}
\affiliation{Instituto de F\'{i}sica Fundamental IFF-CSIC, Calle Serrano 113b, Madrid 28006, Spain}
\author{Paloma A. Huidobro}
\affiliation{Departamento de F\'{i}sica Te\'orica de la Materia Condensada, Universidad Aut\'onoma de Madrid, E-28049 Madrid, Spain}
\affiliation{Condensed Matter Physics Center (IFIMAC), Universidad Aut\'onoma de Madrid, E-28049 Madrid, Spain}
\affiliation{Instituto de Telecomunica\c c\~oes, Instituto Superior Tecnico-University of Lisbon, Avenida Rovisco Pais 1, Lisboa, 1049-001 Portugal}
\author{Jorge Bravo-Abad}
\affiliation{Departamento de F\'{i}sica Te\'orica de la Materia Condensada, Universidad Aut\'onoma de Madrid, E-28049 Madrid, Spain}
\affiliation{Condensed Matter Physics Center (IFIMAC), Universidad Aut\'onoma de Madrid, E-28049 Madrid, Spain}
\author{Alejandro Gonz\'alez-Tudela}
\email{a.gonzalez.tudela@csic.es}
\affiliation{Instituto de F\'{i}sica Fundamental IFF-CSIC, Calle Serrano 113b, Madrid 28006, Spain}

\begin{abstract}

Long-range interactions are a key resource in many quantum phenomena and technologies.
Free-space photons mediate power-law interactions but lack tunability and suffer from decoherence processes due to their omnidirectional emission. Engineered dielectrics can yield tunable and coherent interactions, but typically at the expense
of making them both shorter-ranged and sensitive to material disorder and photon loss.
Here, we propose a platform that can circumvent all these limitations based on three-dimensional subwavelength atomic arrays subjected to magnetic fields. Our key result is to show how to design the polaritonic bands of these atomic metamaterials to feature a pair of frequency-isolated Weyl points. These Weyl excitations can thus mediate interactions that are simultaneously long-range, due to their gapless nature; robust, due to the topological protection of Weyl points; and decoherence-free, due to their subradiant character. We demonstrate the robustness of these isolated Weyl points for a large regime of interatomic distances and magnetic field values and characterize the emergence of their corresponding Fermi arcs surface states. The latter can as well lead to two-dimensional, non-reciprocal atomic interactions with no analogue in other chiral quantum optical setups.

\end{abstract}

\maketitle

Long-range interactions between quantum systems are the source of exotic quantum many-body phenomena~\cite{defenu2023}, such as supersolid~\cite{li2017,leonard2017}, topological~\cite{micheli2006}, and time crystalline~\cite{zhang2017b} phases, or the violation of Lieb-Robinson bounds in the propagation of quantum correlations~\cite{hauke2013,richerme2014}. Besides, they can also be used to build quantum simulators to study frustrated magnetism~\cite{hauke2010,maik2012,bello2022,arguello-luengo2022} or chemistry~\cite{arguello-luengo2019,arguello-luengo2021,malz2023} problems, to generate metrologically-relevant states~\cite{sinatra2022}, or even as a resource for hybrid quantum-classical algorithms~\cite{tabares2023}, among other applications. Unfortunately, finding ``ideal" long-range interactions is still an open challenge in quantum science and technology. The reason is that, ideally, the interactions must not only be long-range, but also decoherence-free, versatile, and robust to other experimental imperfections, like disorder, something hard to achieve simultaneously within the same platform.

Photons are good candidates to mediate long-range interactions between distant quantum emitters thanks to their fast propagation and excellent coherence properties. Already in free-space, they lead to power-law interactions like in Rydberg~\cite{saffman2010} or dipolar atoms~\cite{lahaye2009}. However, their shape cannot be tuned and, more importantly, they are unavoidably accompanied by dissipation due to their eventual propagation into other directions. Engineered dielectrics can suppress to a great extent such dissipation through the use of photonic band-gaps~\cite{joannopoulos_book} and provide some tunability in the interaction form~\cite{gonzalez-tudela2015,douglas2015a,gonzalez-Tudela2018a,perczel2020b,redondo-yuste2021,navarro-baron2021}. However, these advantages generally come at the price of an exponential reduction of the range of the interaction for finite band-gaps~\cite{gonzalez-tudela2015,douglas2015a} and a sensitivity to material disorder. Singular~\cite{gonzalez-Tudela2018a,perczel2020b,redondo-yuste2021,navarro-baron2021} and topological photonic band-gaps~\cite{bello2019,kim2021,leonforte2021a,debernardis2021,ying2019,garcia-elcano2020,garcia-elcano2021} have been shown to provide partial solutions to these two problems resulting in power-law or robust-to-disorder interactions, respectively, in some cases simultaneously, like in three-dimensional Weyl systems~\cite{ying2019,garcia-elcano2020,garcia-elcano2021}. However, all of these platforms face an ultimate challenge, that is, neither singular nor topological bands are protected from photon loss due to the coupling to other propagating photonic channels, which eventually destroy their coherence. Thus, whether it is possible to find a setup that can mediate interactions that are long-range, decoherence-free, and with protection against disorder remains still an open question.

\begin{figure}[t]
    \centering
    \includegraphics[width=\columnwidth]{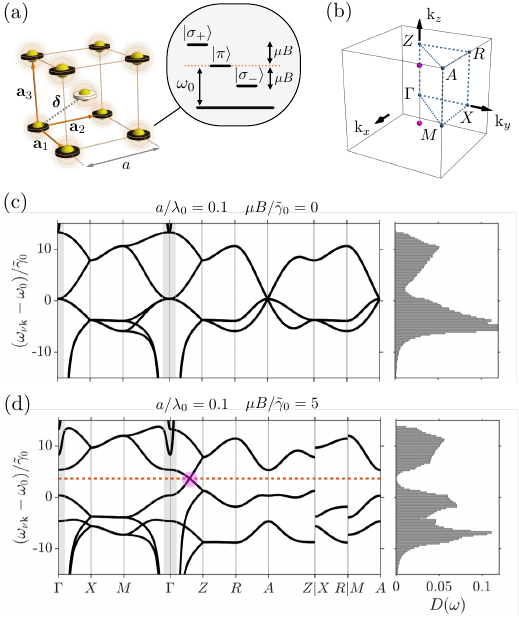}
    \caption{
        (a)~Schematics of the atomic array in real space. The atoms, modeled as 4-level systems, are arranged to form a body-centered-cubic geometry. A uniform magnetic field $B$ is introduced in the $z$ direction.
        (b)~First Brillouin zone associated with the studied lattice. 
        (c)~Band structure along the high-symmetry path and corresponding density of states of an atomic array with $a/\lambda_0=0.1$ in the absence of a magnetic field.
        (d)~Band structure along the high-symmetry path and corresponding density of states of an atomic array with $a/\lambda_0=0.1$ and $\mu B/\tilde{\gamma}_0 = 5$.
        \vspace{-0.3cm}
        }
    \label{fig:Fig1}
\end{figure}

In this Letter, we provide an essential step towards addressing this question proposing a concrete setup where such an ideal scenario can be realized. In particular, it consists of a three-dimensional subwavelength atomic array in a body-centered cubic geometry subjected to a uniform magnetic field [see Fig.~\ref{fig:Fig1}~(a)]. As it occurs in other atomic metamaterials~\cite{bettles2016,shahmoon2017,facchinetti2016,asenjo-garcia2017,syzranov2016,bettles2017,perczel2017a,perczel2017b}, this system hosts long-lived (subradiant) polaritonic excitations that can mediate interactions between additional impurity atoms whose form depends on the array's dispersion~\cite{masson2020,patti2021,brechtelsbauer2021,fernandez-fernandez2022}. The key result of this work is to demonstrate that, under certain conditions, the bulk subradiant excitations of the considered atomic arrays [see Fig.~\ref{fig:Fig1}~(a)] feature frequency-isolated Weyl points, unlike other similar proposals~\cite{syzranov2016}. This means that they can mediate robust, power-law photon-mediated interactions, as predicted in Refs.~\cite{ying2019,garcia-elcano2020,garcia-elcano2021} for simplified photonic lattices, but with the added advantage of the isolation from the rest of the propagating modes, that will make the interactions virtually decoherence-free. Besides, using exact diagonalization in slab geometries, we also demonstrate the emergence of strongly subradiant Fermi arc surface states with a notable polarization texture. The latter opens the path to obtain non-reciprocal photon-mediated interactions~\cite{garcia-elcano2023}, typical of chiral quantum optical setups~\cite{lodahl2017}.

\noindent \textit{Setup.-} We consider a collection of atomic emitters pinned to the positions of a body-centered cubic lattice. In Fig.~\ref{fig:Fig1}(a), we plot the real space representation of the array that we describe as a non-Bravais lattice, with the orange arrows indicating the direct lattice vectors. The unit cell encompasses two sites whose relative position is given by the sublattice displacement $\bm{\delta}=a(\mathbf{e}_x+\mathbf{e}_y+\mathbf{e}_z)/2$, where $a$ stands for the lattice parameter and $\mathbf{e}_{x,y,z}$ denote the three Cartesian unit vectors. We assume that all atoms are identical and focus on a $J=0\rightarrow J=1$ optical transition, i.e., we model each atom as a four-level system, with a singlet ground state $\ket{g}$ and a triply degenerated excited state $\{\ket{\sigma_{\pm}},\ket{\sigma_{z}}\}$. The frequency difference between the ground and excited states is given by $\omega_0=2\pi c/\lambda_0=c k_0$, where $\lambda_0$ ($k_0$) denotes the transition's wavelength (momentum) and $c$ the speed of light in vacuum. This scenario can be attained, e.g., by harnessing the $^3P_0$ $-$ $ ^3D_1$ transition in alkaline-earth-metal atoms~\cite{olmos2013}.

\noindent \textit{Bulk results.-} By working in the Markovian approximation (which is well justified in these atomic setups since  $\gamma_0\ll\omega_0$, where $\gamma_0$ is the single atom decay rate) and performing an adiabatic elimination of the photonic degrees of freedom, one can define an effective dipole-dipole Hamiltonian $\hat{H}_\text{eff}$ for the system's atomic wavefunction~\cite{asenjo-garcia2017}. In the limit of very large systems, we can assume periodic boundary conditions and diagonalize $\hat{H}_\text{eff}$ exactly in the single-excitation sector introducing Bloch states $\ket{\psi_{\nu\kk}}$. Doing this diagonalization and setting $\hbar=1$, one finds that (see Supplemental Material~\cite{suppl}):
\begin{equation}
    \hat{H}_\mathrm{eff} = \sum_{\nu,\kk}
    \left(\omega_{\nu\kk}-i\frac{\gamma_{\nu\kk}}{2}\right)
    \ket{\psi_{\nu\kk}}\bra{\psi_{\nu\kk}},
\end{equation}
where $\omega_{\nu\kk}$ and $\gamma_{\nu\kk}$ account for the dispersive and dissipative components of the atomic array's dynamics, respectively. Here, the index $\nu$ runs over bands, whereas the quasimomentum $\kk$ denotes a vector belonging to the first Brillouin zone which, for the considered geometry, takes the form of the cube of length $2\pi/a$ depicted in Fig.~\ref{fig:Fig1}(b). 
The Bloch states with $\kk$ values fulfilling that $|\kk|\geq \omega_{\nu\kk}/c$  belong to the so-called light cone region. In 1D and 2D periodic arrays, the modes characterized by quasimomenta lying inside that region are generally leaky whereas, outside the light cone, they feature a vanishing dissipative component owing to the energy and momentum mismatch with the free-space photons~\cite{asenjo-garcia2017}. However, the eigenvalues of $\hat{H}_\text{eff}$ obtained for 3D periodic lattices are purely real irrespective of whether they lie inside or outside the light cone~\cite{brechtelsbauer2021}. This is because, in that case, the atomic ensemble occupies the whole physical space so that lattice excitations have no place to escape but to the lattice itself. The lack of dissipation in the infinite 3D lattice allows us to focus on the dispersive properties of the atomic array and, more precisely, on how we can obtain a Weyl energy dispersion. Finally, let us note that we always take into account all near-, mid-, and far-field contributions of the dipole-dipole interactions in the calculations, thus allowing us to capture the strong modification of the bands close to the border of the light cone (gray shadowed area)~\cite{blancodepaz2022}. This is relevant to characterize faithfully the robustness of the spectral features of this system. 

\begin{figure}[t]
    \centering
    \includegraphics[width=8.7cm]{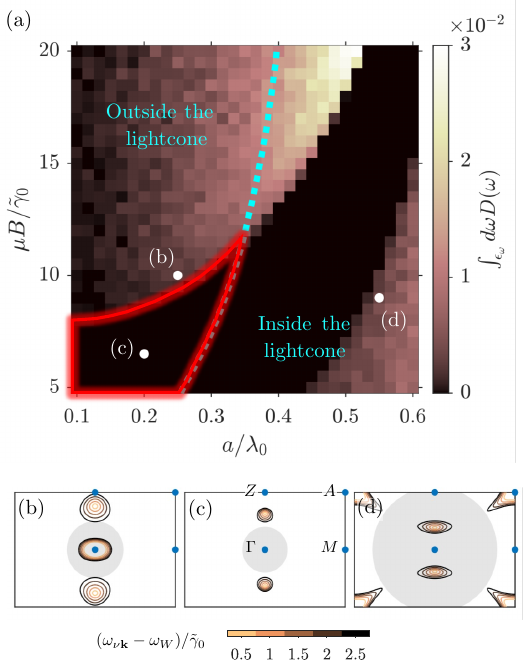}
    \caption{
        (a)~Density of states evaluated at the Weyl frequency as a function of the magnetic field and the interatomic distance. The color code measures the number of states in an energy range of $\Delta\omega=0.1\tilde{\gamma}_0$ around the Weyl frequency. The cyan line divides the phase space into two halves: the left one (smaller interatomic distances) corresponds to configurations for which the Weyl points lie outside the light cone; the right one (larger interatomic distances) corresponds to configurations for which the Weyl points lies inside the light cone.
        (b-c)~Equifrequency contours projected along the $k_x-k_y=0$ plane of the first Brillouin zone for values slightly above the Weyl frequency in different points of the phase diagram. Weyl points emerge in the $k_z$ axis. The only case in which the Weyl points are frequency isolated is in (c). In (b) and (d) frequency isolation is spoiled by the incursion of parabolic bands centered in the $\Gamma$ and $A$ points, respectively. The gray shadow circle represents the light cone.
        \vspace{-0.3cm}
        }
	\label{fig:Fig2}
\end{figure}

Let us first illustrate the band structure of the atomic array without adding the magnetic field. This is what we show in Fig.~\ref{fig:Fig2}(c) for an interatomic distance $a/\lambda_0=0.1$. In the left panel, we plot $\omega_{\nu\kk}$ centered around $\omega_0$ and in units of $\tilde{\gamma}_0=\gamma_0/(k_0 a)^3$, that is the energy scale of the dipole-dipole interactions in the deep subwavelength limit. Besides, in the right panel, we plot the associated density of states $D(\omega)$. The main observation of Fig.~\ref{fig:Fig1}(c) is that around $\omega_0$ the system features a three-fold degeneracy at $\Gamma$ and three doubly degenerate bands that merge at the $A$ point, but no clear signature of Weyl points. This is reflected in the density of states, which features a dip, but retaining a finite value, around $\omega_0$.

The absence of Weyl points in the previous configuration is expected since one needs to break either inversion (I) or time-reversal (T) symmetry to find the Weyl phase~\cite{lu2013}. Interestingly, the latter can be done in these atomic metamaterials, e.g., by introducing a uniform magnetic field along the positive $z$ direction which lifts the degeneracy of the triplet state by the Zeeman splitting $\mu B$ [see Fig.~\ref{fig:Fig1}(a)] and thus breaks T~\cite{footnote1}. In Fig.~\ref{fig:Fig1}(d), we plot the new band structure and density of states for the same interatomic distance as in panel (c), but for a finite magnetic field $\mu B/\tilde{\gamma}_0=5$. There, we observe how the magnetic field lifts the degeneracies at $\Gamma$ and $A$, and, consequently, an unavoidable crossing, that we identify as a Weyl point, appears in the $\Gamma Z$ segment. Importantly, this two-band singularity dwells in a complete frequency gap, which translates into a clean parabolic dependence of $D(\omega)$ in the surroundings of that frequency value [see right panel of Fig.~\ref{fig:Fig1}(d)], as it is expected in the presence of frequency-isolated Weyl singularities. The latter contrasts with the Weyl dispersion emerging in other simpler geometries~\cite{syzranov2016} for which such frequency isolation cannot be achieved (see Supplemental Material~\cite{suppl}), and thus are unsuitable to mediate the perfectly coherent interactions between impurity atoms.

Let us now study more in detail the robustness of the frequency isolation of the Weyl points of our setup.  For that, we examine a large set of configurations spanning different $a/\lambda_0$ and $\mu B/\tilde{\gamma}_0$ ratios and use the vanishing density of states around the Weyl frequency $\omega_W$ as a figure of merit of its stability. These calculations result in the stability diagram of Fig.~\ref{fig:Fig2}(a). For all the considered configurations, the system harbors a pair of Weyl singularities at frequency $\omega_W$. The color scale accounts for the density of states of the atomic ensemble integrated over a small frequency range around $\omega_W$. Therefore, we can say that the Weyl points remain frequency-isolated in the black central region, where the density of states at $\omega_W$ essentially vanishes, while its isolation is spoiled in the colored areas. For small $a/\lambda_0$ and large $\mu B/\tilde{\gamma}_0$ (upper left corner of the phase diagram), the latter occurs because a parabolic band centered at $\Gamma$ intercepts the Weyl frequency. For large $a/\lambda_0$ and small $\mu B/\tilde{\gamma}_0$ (lower right corner of the phase diagram) we observe a similar situation but, there, the parabolic band intercepting the Weyl frequency has its minimum at $A$. We illustrate these two cases in Figs.~\ref{fig:Fig2}(b) and (d), respectively, by plotting the projection over the $k_x-k_y=0$ plane of the equifrequency contours associated with a set of frequency values slightly above $\omega_W$. For comparison, we also display an example wherein the Weyl points are frequency isolated in Fig.~\ref{fig:Fig2}(c).

Another observation is that by changing the magnitude of the applied magnetic field we can modify the positions of the Weyl points which, owing to the lattice symmetries, are restricted to move along the $k_z$ axis (see Supplemental Material~\cite{suppl}). At the same time, the extension of the light cone region is determined by the $a/\lambda_0$ ratio. In particular, it occupies more of the first Brillouin zone as the interatomic separation increases for fixed $\lambda_0$. Consequently, and irrespective of how the Weyl points move in reciprocal space, we find that, for every Zeeman splitting value, there is some critical $a/\lambda_0$ ratio up to which the Weyl points fall inevitably into the light cone region. We mark this frontier using a cyan dotted line that divides the parameters' space into two halves, left and right, depending on whether the Weyl points lie outside or inside the light cone, respectively. This observation permits us to identify a region in the lower left corner of the phase diagram where the Weyl points are frequency-isolated and live outside the light cone. As we will see next, these will have important consequences when finite-size effects are considered.

\begin{figure}[t]
    \centering
    \includegraphics[width=8.7cm]{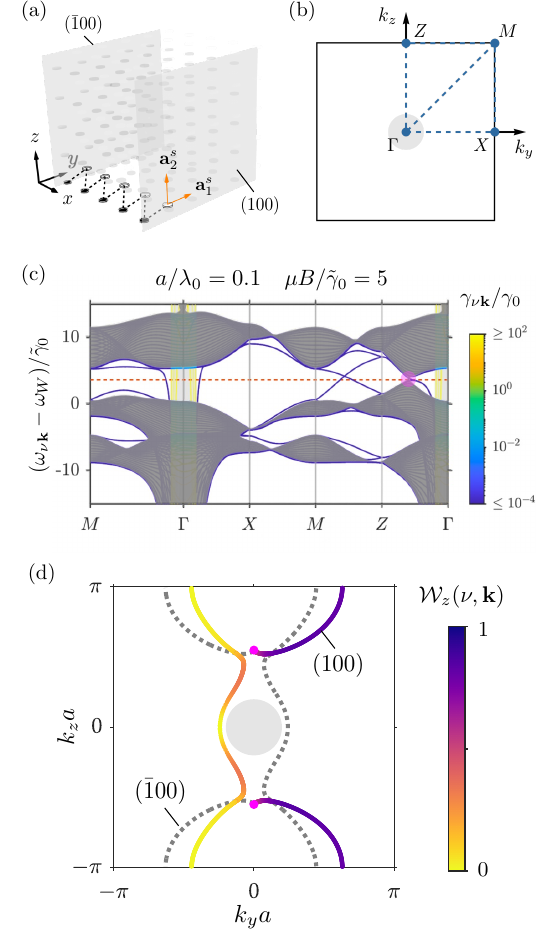}
    \caption{
        (a)~Real space sketch of the slab geometry provided that the lattice is cut along the $x$ direction. 
        (b)~Surface Brillouin zone associated with the considered cut. The gray disk represents the light cone region corresponding to the configuration for which $a/\lambda_0=0.1$.
        (c)~Slab bands along a high-symmetry path for the studied configuration. The color legend accounts for the dissipative component $\gamma_{\nu\kk}$.
        (d)~Equifrequency contours of the slab bands at the Weyl frequency. The colored lines account for the modes localized in the $(100)$ facet, whereas the gray dashed lines display the modes associated with the $(\bar{1}00)$ facet. The color code indicates the total weight of the $z$-polatization in each Bloch mode comprising the arcs. The magenta dots mark the projection of the Weyl points over the surface Brillouin zone. 
        \vspace{-0.3cm}
        } \label{fig:Fig3}
   \end{figure}

\noindent \textit{Subradiant surface excitations.-} One of the hallmarks of Weyl systems, which is also a clear signature of their topological nature, is that they support robust surface states -- usually referred to as Fermi arcs in the electronic context~\cite{wan2011a}. To see whether they also appear in our system and characterize their stability, we implement a slab geometry for the atomic array by cutting the lattice in a given direction while maintaining the periodicity of the array in the other two, see Fig.~\ref{fig:Fig3}(a). The resulting problem is equivalent to an infinite 2D lattice with a unit cell extending in the out-of-plane direction such that, as we increase the width of the slab, we approach the infinite 3D limit in which we have focused so far. A key difference between the two configurations, though, is that the slab array no longer occupies the whole physical space, and thus their excitations will acquire a finite dissipative component $\gamma_{\nu\kk}$. However, we expect this dissipative component to be very small ($\gamma_{\nu\kk}\ll \gamma_0$) in the region where the slab bands lie outside the light cone.

To certify these expectations, we consider a slab geometry with a finite width along $x$, featuring two boundaries that we identify as the $(100)$ and the $(\bar{1}00)$ facets, as schematically depicted in Fig.~\ref{fig:Fig3}(a). There, the unit cell is represented by the collection of black and white shallow cylinders connected by a dashed line. Its length determines the width of the slab that we take in the following to be $w/a=15.5$ (corresponding to a unit cell composed of 30 non-equivalent sites). The gray planes mark the slab terminations and the orange arrows denote the surface lattice vectors $\mathbf{a}_1^{s}$ and $\mathbf{a}_2^{s}$ that coincide with the $y$ and $z$ directions. The surface Brillouin zone associated with the chosen cut, which takes the form of a square of size $2\pi/a$, is plotted in Fig.~\ref{fig:Fig3}(b). The small circle inscribed in the square represents the light cone region corresponding to the $a/\lambda_0=0.1$ case. 

In Fig.~\ref{fig:Fig3}(c) we plot the band structure of the atomic slab for $a/\lambda_0=0.1$ and $\mu B/\tilde{\gamma}_0=5$, where the dark gray cover accounts for the bulk bands projected along the cut direction. At the Weyl frequency (red dotted line), the bulk bands touch at some point in the $\Gamma Z$ segment that, in turn, coincides with the projection of the Weyl point over the surface Brillouin zone (magenta circle). Nonetheless, the interesting bands here are those colored, which are the ones emerging because of the finite width of the slab. There, we can identify a set of superradiant bands within the light cone (in yellow). These are related to the hybridization of the different layers of the slab due to the long-range nature of the free-space interactions. Even though they intersect the Weyl frequency, their contribution to the density of states is small due to the high group velocity, and thus we expect them to play a spurious role when an impurity atom is resonant with the Weyl frequency. Outside of the light cone, we observe a set of subradiant bands some of which also cross the Weyl frequency. The modes comprising those bands are precisely the ones that will form the polaritonic Fermi arcs.

To make the connection with Fermi arcs more evident, we plot in Fig.~\ref{fig:Fig3}(d) the equifrequency contours of the subradiant slab bands at the Weyl frequency. There, the arc shape of the Weyl surface states becomes apparent. By examining the localization of the eigenstates that form the equifrequency lines (see Supplemental Material~\cite{suppl}), we can determine the facet to which each of the modes belongs. The curves corresponding to the set of modes localized in the $(100)$ and the $(\bar{1}00)$ boundaries are shown using colored solid and gray dashed lines, respectively. In what follows, we focus our discussion on the surface modes along the $(100)$ surface, though similar statements can be made for the other.  

As we observe in Fig.~\ref{fig:Fig3}(d), an open arc connecting the projection of the Weyl singularities (magenta dots) appears in the right half of the surface Brillouin zone. Together with this open arc, we find a second contour line forming a closed curve over the torus geometry of the surface Brillouin zone. Similarly to what is reported in photonic Weyl metacrystals~\cite{Yang2018}, these arcs form part of the same helicoidal structure that defines the dispersion of the edge modes. 

Finally, we also study the polarization texture of the states comprising the equifrequency lines associated with the surface states in the $(100)$ facet. The color code in Fig.~\ref{fig:Fig3}(d) shows the weight of the $z$-polarization component $\mathcal{W}_z(\nu,\kk)$ for each mode in the considered equifrequency contours (see Supplemental Material~\cite{suppl}). As seen, the modes defining the open arcs connecting the Weyl points projections feature a strong $z$-polarized flavor. Conversely, the main contributions to the ones forming the close curve correspond to the $x$ and $y$ polarizations. This result shows that the subradiant topological surface modes of a Weyl atomic array display a marked polarization texture, thus, opening new avenues for engineering chiral quantum optical scenarios~\cite{lodahl2017} beyond one-dimension~\cite{garcia-elcano2023}.

\noindent \textit{Conclusions \& Outlook.-} Summing up, we demonstrate that it is possible to obtain frequency-isolated Weyl energy dispersions in the subradiant excitations of subwavelength atomic arrays. By extensive numerical analysis, we analyze their robustness as a function of relevant experimental parameters such as the magnetic field, interatomic distances, and the slab width, identifying the optimal windows to obtain them. Besides, we also demonstrate the emergence of strongly subradiant, chiral surface states associated with the Fermi arc modes. Combined with previous works~\cite{masson2020,patti2021,brechtelsbauer2021,fernandez-fernandez2022}, our results open an avenue to engineer decoherence-free, robust, and long-range interactions between additional impurity atoms when harnessing their bulk excitations~\cite{ying2019,garcia-elcano2020,garcia-elcano2021} or non-reciprocal, dissipative interactions~\cite{garcia-elcano2023} when considering their surface modes.

There are several interesting perspectives opened by our work. The most obvious one is to take our theoretical proposal into practice, e.g., using setups similar to the first implementations of such subwavelength atomic arrays~\cite{rui2020,srakaew2023}. On the theoretical front, it will be interesting to look into the non-Hermitian topological effects~\cite{bergholtz2021}, something scarcely explored in three-dimensions, or to study the effect of the intrinsic non-linearity of their constituents~\cite{rusconi2021}.

\acknowledgments{AGT and IGE acknowledge support from the Proyecto Sin\'ergico CAM 2020 Y2020/TCS-6545 (NanoQuCo-CM), the CSIC Research Platform on Quantum Technologies PTI-001 and from Spanish projects PID2021-127968NB-I00 funded by MICIU/AEI/10.13039/501100011033/ and by FEDER Una manera de hacer Europa, and TED2021-130552BC22 funded by MICIU/AEI /10.13039/501100011033 and by the European Union NextGenerationEU/ PRTR, respectively. AGT also acknowledges a 2022 Leonardo Grant for Researchers and Cultural Creators, and BBVA Foundation. 
PAH and JBA acknowledge funding through the “María de Maeztu” Programme for Units of Excellence in R\&D (CEX2023-001316-M).
PAH acknowledges funding from the Spanish Ministry for Science and Innovation (Grant No. RYC2021-031568-I and project No. PID2022-141036NA-I00 financed by MCIN/AEI/10.13039/501100011033 and FSE+) and the Fundaçao para la Cîencia e a Tecnologia under project 2022.06797.PTDC. 
JBA acknowledges support from grant PID2022-139995NB-I00 funded by MICIU/AEI/ 10.13039/501100011033.
}


%

\widetext
\clearpage
\begin{center}
\textbf{\large Supplemental Materials: Long-range interactions in Weyl dense atomic arrays protected from dissipation and disorder}
\end{center}

\setcounter{equation}{0}
\setcounter{figure}{0}
\setcounter{table}{0}
\setcounter{page}{1}
\makeatletter
\renewcommand{\thesection}{S\arabic{section}}
\renewcommand{\theequation}{S\arabic{equation}}
\renewcommand{\thefigure}{S\arabic{figure}}
\renewcommand{\bibnumfmt}[1]{[S#1]}
\renewcommand{\citenumfont}[1]{S#1}

In this supplementary material, we provide more details on how to obtain the results of the main text. In particular:
\begin{itemize}
    \item In Section~\ref{sec:Hamiltonian}, we explain how to obtain the effective spin Hamiltonian describing the atomic array, and how to diagonalize it in the single excitation subspace.
    \item In Section~\ref{sec:comparison}, we study the polaritonic bands of the simple cubic geometry to evidence the absence of Weyl points that are frequency isolated.
    \item In Section~\ref{sec:movement}, we explain the movement of the Weyl points and their position with respect to the light cone as we vary both the interatomic distances and magnetic fields.
    \item Finally, in Section~\ref{sec:surface}, we investigate the confinement of the surface modes corresponding to the Fermi arcs.    
\end{itemize}

\section{Light-matter Hamiltonian for an atomic lattice~\label{sec:Hamiltonian}}

\subsection{Dipolar Hamiltonian for an atomic ensemble}

The system under consideration consists of a collection of atoms tightly trapped at the nodes of an optical lattice. We assume that each lattice site harbors a single particle that we treat as a hydrogenic atom. The atoms are coupled to the free-space modes of the electromagnetic field by an electric-dipole-transition connecting a non-degenerate singlet ground state, with well-defined angular momentum $J=0$, and a triply degenerated excited state with $J=1$. 
In other words, the atoms are modeled as four-level systems with each of their internal states being characterized by their principal and secondary angular momentum quantum numbers: $J$ and $m_J$, respectively. The total Hilbert space of the atomic part is spanned by the set of states:
\begin{align}
    & \ket{g}_j\phantom{_+}\coloneqq\ket{J=0,\,m_J=0}_j
    \\[1.0ex]
    & \ket{\sigma_\pm}_j\coloneqq\ket{J=1,\,m_J=\pm1}_j
    \\[1.0ex]
    & \ket{\pi}_j\phantom{_+}\coloneqq\ket{J=1,\,m_J=0}_j
\end{align}
where the index $j$ labels the atoms in the lattice arrangement.

The light-matter Hamiltonian describing the coupling of the atomic array with the vacuum modes under the dipole approximation reads (setting $\hbar=1$)
\begin{equation}
    \hat{H}=
    \sum_j\left[\sum_{\alpha}^{\sigma_{\pm},\pi}
    \omega_{j}\ket{\alpha}_j\bra{\alpha}_j
    -\hat{\bm{\dip}}_j\,\hat{\mathbf{E}}_\perp(\rr_j)\right]
    +\sum_{\kkk}\sum_{\eta} \omega_\kkk\,\hat{a}^\dagger_{\kkk\eta}\hat{a}_{\kkk\eta}.
\end{equation}
Here, $\kkk$ stands for the wave vector of the free photonic excitation, whose associated dispersion is $\omega_\kkk=|\kkk|c$, and $\hat{a}_{\kkk\eta}$ represents the usual bosonic annihilation operator, with $\eta$ labeling the polarization degrees of freedom. 
From now on, we assume that all atoms are identical which allows us to replace the $j$-dependent atomic transition frequency $\omega_j$ and dipole operator $\hat{\bm{\dip}}_j$ by $\omega_0$ and $\hat{\bm{\dip}}_0$, respectively. The specific values of these quantities will depend on the considered atomic specie.
Finally, we write $\hat{\mathbf{E}}_\perp(\rr_j)$ as the transverse electric field operator evaluated at the $j$-th atomic position: 
\begin{equation}
    \hat{\mathbf{E}}_\perp(\rr_j)=\sum_{\kkk}\sum_{\eta}
    \left[
    i\sqrt{\frac{\omega_\kkk}{2\varepsilon_0V}}\,\bm{\epsilon}_\eta\,e^{i\kkk\rr_j}\,\hat{a}_{\kkk\eta}+\mathrm{H.c}
    \right],
\end{equation}
where $\bm{\epsilon}_\eta$ denotes the polarization vector accounting for the two possible components of the transversal electric field (note that $\bm{\epsilon_\eta}\perp\kkk$ for $\eta=1,2$), $\varepsilon_0$ is the vacuum permittivity, and $V$ stands for the quantization volume. 
Interestingly, the energy degeneracy of the triplet state, which reflects the isotropic nature of the atomic transition~\cite{james1993}, can be lifted by the introduction of a uniform magnetic field. This produces a Zeeman splitting of the energy levels:
\begin{equation}
    \hat{H}_\mathrm{Zeeman}=\bm{\mu} \mathbf{B}
    \left(
    \ket{\sigma_+}_j\bra{\sigma_+}_j-\ket{\sigma_-}_j\bra{\sigma_-}_j
    \right),
\end{equation}
where $\bm{\mu}\mathbf{B}$ is the Zeeman shift of the atomic levels with magnetic momentum $\bm{\mu}$ under the magnetic field $\mathbf{B}$.
To facilitate the evaluation of the dipolar term it is convenient to introduce the Cartesian basis which results from the Cartesian decomposition of the radial unit vector. Without loss of generality, we choose the $z$ direction as the quantization axis and define $\mathbf{B}=B\,\mathbf{e}_z$. Then, using that $\ket{\sigma_\pm}_j=\mp\frac{1}{\sqrt{2}}(\ket{x}_j\pm i\ket{y}_j)$ and $\ket{\pi}_j=\ket{z}_j$, the dipole Hamiltonian can be rewritten as follows
\begin{equation}
    \hat{H}=
    \sum_{j}\left[
    \omega_{0}\,\hat{\mathbf{b}}_j^\dagger\hat{\mathbf{b}}_j
    -\hat{\bm{\dip}}_0\,\hat{\mathbf{E}}_\perp(\rr_j)
    +\hat{H}_\mathrm{Zeeman}
    \right]
    +\sum_{\kkk}\sum_{\eta} \omega_\kkk\,\hat{a}^\dagger_{\kkk\eta}\hat{a}_{\kkk\eta},
\end{equation}
where $\hat{\mathbf{b}}_j=\sum_\beta^{x,y,z}\hat{b}_{j\beta}\mathbf{e}_\beta$ and $\hat{\mathbf{b}}_j^\dagger=\sum_\beta^{x,y,z}\hat{b}_{j\beta}^\dagger\mathbf{e}_\beta$ (with $\hat{b}_{j\beta}=\ket{g}_j\bra{\beta}_j$ and $\hat{b}_{j\beta}^\dagger=\ket{\beta}_j\bra{g}_j$) define the vector transition operators for the $j$-th atom~\cite{james1993,antezza2009a,olmos2013}. Using this notation, the dipole operator can be expressed as $\hat{\bm{\dip}}_0=\dip_0\,(\hat{\mathbf{b}}_j^\dagger+\hat{\mathbf{b}}_j)$, where $\dip_0$ accounts for the corresponding matrix element. Note that, in the new basis, the Zeeman term is given by 
\begin{equation}
    \hat{H}_\mathrm{Zeeman}=-i\mu B\left(\ket{x}_j\bra{y}_j-\ket{y}_j\bra{x}_j\right),
\end{equation}
with $\mu$ being the projection of $\bm{\mu}$ over the direction of the magnetic field.

\subsection{Effective Hamiltonian for the atomic degrees of freedom}

By invoking the Born-Markov approximation, one can perform an adiabatic elimination of the photonic degrees of freedom which results in the formulation of a master equation for the reduced density matrix $\hat{\rho}_s(t)$ describing only the atomic degrees of freedom~\cite{james1993,olmos2013}.
Alternatively, the dynamics of the atomic ensemble can be described in the quantum jump formalism of open quantum systems via the definition of a non-Hermitian effective Hamiltonian $H_\mathrm{eff}$~\cite{asenjo-garcia2017,perczel2017b}. More precisely, one has that the temporal evolution of $\hat{\rho}_s(t)$ follows the form
\begin{equation}
    \partial_t\hat{\rho}_s(t)=
    -i\left[\hat{H}_\mathrm{eff}\hat{\rho}_s(t)-\hat{\rho}_s(t) \hat{H}_\mathrm{eff}^\dagger\right] + \hat{\Upsilon}.
\end{equation}
Here, $\hat{\Upsilon}$ denotes the so-called recycling operator that accounts for stochastic quantum jumps occurring among the ground state levels of the emitters and is explicitly given by
\begin{equation}
    \hat{\Upsilon}=
    \frac{6\pi\gamma_0}{k_0}\,
    \sum_{jj^\prime}\sum_{\beta\beta^\prime}^{x,y,z}
    \mathrm{Im}\,G_{\beta\beta^\prime}(\rr_j,\rr_{j^\prime},k_0)
    \ket{g}_j\bra{\beta}_j
    \rho_s
    \ket{\beta^\prime}_{j^\prime}\bra{g}_{j^\prime}.
\end{equation}
where $\gamma_0=\bm{\dip}_0^2 k_0^3/(3\pi\varepsilon_0)$, with $k_0=\omega_0/c$, is the spontaneous emission rate for a single atom in free-space and
\begin{align}
\label{eq:GreensFunMatElem}
\begin{split}
    G_{\beta\beta^\prime}(\rr_j,\rr_{j^\prime},k_0) =
    & \frac{e^{i k_0|\rr_{jj^\prime}|}}{4\pi|\rr_{jj^\prime}|}
    \left[
    \left( 1+\frac{i}{k_0|\rr_{jj^\prime}|}-\frac{1}{(k_0|\rr_{jj^\prime}|)^2} \right)
    \delta_{\beta\beta^\prime}
    \right.
    \\[1.0ex]
    & \left.
    +\left(-1-\frac{3i}{k_0|\rr_{jj^\prime}|}+\frac{3}{(k_0|\rr_{jj^\prime}|)^2} \right)
    \frac{[\rr_{jj^\prime}]_\beta\,[\rr_{jj^\prime}]_{\beta^\prime}}{|\rr_{jj^\prime}|^2}
    \right]
    - \frac{\delta_{\beta\beta^\prime}\,\delta^{(3)}(\rr_{jj^\prime})}{3k_0^2},
\end{split}
\end{align}
with $\rr_{jj^\prime}\equiv\rr_j-\rr_{j^\prime}$, stand for the matrix elements of the electromagnetic dyadic Green's function in free-space~\cite{novotny_book, buhmann_book}. 
However, provided that we restrict ourselves to the single excitation sector, the contribution to the dynamics by $\hat{\Upsilon}$ can be safely neglected~\cite{perczel2017b}. The temporal evolution of the atomic wavefunction is thus completely determined by the effective Hamiltonian $\hat{H}_\mathrm{eff}$, which reads
\begin{equation}
\label{eq:H_eff}
    \hat{H}_\mathrm{eff} = 
    \sum_{j}\sum_{\beta\beta^\prime}^{x,y,z}
    \left[
    \Omega_{\beta\beta^\prime} \,\hat{b}^\dagger_{j\beta}\hat{b}_{j\beta^\prime}
    -\frac{3\pi\gamma_0}{k_0}\,\sum_{j^\prime\neq j}
    G_{\beta\beta^\prime}(\rr_j,\rr_{j^\prime},k_0) \,\hat{b}^\dagger_{j\beta}\hat{b}_{j^\prime\beta^\prime}
    \right],
\end{equation}
with
\begin{equation}
    \Omega_{\beta\beta^\prime}=
    \begin{cases}
        \omega_0-i\frac{\gamma_0}{2}  &\text{if }\beta=\beta^\prime,\\
        - i\mu B                      &\text{if }\beta=x,\,\beta^\prime=y,\\
        + i\mu B                      &\text{if }\beta=y,\,\beta^\prime=x,\\
        0                             & \text{otherwise}.
    \end{cases}
\end{equation}

\subsection{The infinite atomic lattice}

Now, we address the situation in which the emitters form an ordered atomic ensemble of infinite size. In that case, the Bloch theorem can be invoked to diagonalize the atomic array in the single excitation sector. The simplest scenario assumes that the atoms form a Bravais lattice $\Lambda$ characterized by the set of points $\RR=\sum_in_i\,\mathbf{a}_i$, where $n_i\in\mathbb{Z}$ and $\mathbf{a}_{i}$ define the direct lattice vectors. However, the most general case corresponds to having several non-equivalent atoms in the unit cell, thereby motivating the introduction of the sublattice degree of freedom $\xi$. For both Bravais and non-Bravais lattices, the system is invariant under discrete translations of the type $\rr\rightarrow\rr+\RR$.

To clarify the upcoming treatment, we make the following notation's change for the atomic operators: $\hat{b}_{j\beta}^{(\dagger)}\rightarrow\hat{b}_{\RR\xi\beta}^{(\dagger)}$. This notation, in turn, evidences the pseudo-spin character of the sublattice degree of freedom.
Then, we define the sublattice displacement $\bm{\delta}_\xi$, that indicates the position of the atom belonging to sublattice $\xi$ inside the unit cell, and introduce the single-particle operator
\begin{equation}
    b_{\kk\xi\beta}^\dagger =
    \frac{1}{\sqrt{N_\xi}}\sum_{\RR}e^{i\kk(\RR+\bm{\delta}_\xi)}b^\dagger_{\RR\xi\beta},
\end{equation}
where $N_\xi$ is the number of sites belonging to the $\xi$-th sublattice and $\kk$ must be understood here as a quasimomentum quantum number. By inverting $b_{\kk\xi\beta}^\dagger$ and $b_{\kk\xi\beta}$ (given by the hermitian conjugate of the former one) and substituting in Eq.~\eqref{eq:H_eff}, $\hat{H}_\text{eff}$ can be re-expressed in a quadratic form, such that:
\begin{equation}
\hat{H}_\mathrm{eff}=\sum_{\kk}\hat{\mathbf{B}}_\kk^\dagger\,H_\text{eff}(\kk)\,\hat{\mathbf{B}}_\kk,
\end{equation}
where $\mathbf{B}_\kk^\dagger$ and $\mathbf{B}_\kk$ stand for row and column operators spanning the orbital and sublattice degrees of freedom. More precisely, we chose the following ordering for the atomic array operators: 
\begin{equation}
    \mathbf{B}_\kk^\dagger=
    [\;
    \underbrace{b_{\kk\xi_1x}^\dagger,\;b_{\kk\xi_1y}^\dagger,\;b_{\kk\xi_1z}^\dagger,}_\text{1st sublattice}
    \quad\dots\quad
    \underbrace{b_{\kk\xi_i x}^\dagger,\;b_{\kk\xi_i y}^\dagger,\;b_{\kk\xi_i z}^\dagger,}_\text{$i$-th sublattice}
    \quad\dots\quad
    \underbrace{b_{\kk \xi_M x}^\dagger,\;b_{\kk \xi_M y}^\dagger,\;b_{\kk \xi_M z}^\dagger}_\text{$M$-th sublattice}
    ].
\end{equation}
Here, we enumerate the different sublattices with the Latin letter $i$, ranging from $1$ to $M$, i.e., we consider $M$ sublattices. This ordering determines the shape of the Bloch Hamiltonian $H_\text{eff}(\kk)$ which is a square matrix of dimension $\mathcal{N}=3\times M$, where the factor of $3$ arises due to the three possible polarizations. 

For the sake of clarity, we define $H_\text{eff}(\kk)$ as the sum of two contributions coming from the first and second terms in Eq.~\eqref{eq:H_eff}, respectively: 
\begin{equation}
    H_\mathrm{eff}(\kk)=H_\text{eff,I}+H_\text{eff,II}(\kk)
\end{equation}
The first contribution $H_\text{eff,I}$ is independent of $\kk$ and consists of a block diagonal matrix featuring the following form
\begin{equation}
    H_\text{eff,I}=
    \begin{bmatrix}
    h_\mathrm{I}^{(1)}  & \cdots & 0
    \\[0.5ex]
    \vdots             & \ddots & \vdots
    \\[0.5ex]
    0                   & \cdots & h_\mathrm{I}^{(M)}
    \end{bmatrix},
\end{equation}
It is worth noting that, since we have assumed a single atomic specie, all the constituent blocks $h_\mathrm{I}^{(i)}$ are identical, i.e., $h_\mathrm{I}^{(i)}=h_\mathrm{I}$ $\forall$ $i$. Each of these blocks is given by a $3\times3$ matrix that reads
\begin{equation}
    h_\mathrm{I}=
    (\omega_0-i\frac{\gamma_0}{2})\mathbb{I}+h_\mathrm{Zeeman},
\end{equation}
with
\begin{equation}
    h_\mathrm{Zeeman}=i\mu B
    \begin{bmatrix}
    0 & -1 & 0
    \\[0.1ex]
    +1 & 0 & 0
    \\[0.1ex]
    0 & 0 & 0
    \end{bmatrix}.
\end{equation}
The second contribution $\bar{H}_\text{eff,II}(\kk)$ contains the dependence on $\kk$ and features a more complicated structure than the previous one, namely
\begin{equation}
    H_\text{eff,II}(\kk)=
    \begin{bmatrix}
    h_\mathrm{II}^{(\xi_1\xi_1)}   & \cdots  &                          & \cdots  & h_\mathrm{II}^{(\xi_1\xi_M)}
    \\[0.5ex]
    \vdots                 & \ddots  &                          & \iddots & \vdots
    \\[0.5ex]
                           &         & h_\mathrm{II}^{(\xi_i\xi_i)} &         & 
    \\[0.5ex]
    \vdots                 & \iddots &                          & \ddots  & \vdots
    \\[0.5ex]
    h_\mathrm{II}^{(\xi_M\xi_1)} & \cdots  &                          & \cdots  & h_\mathrm{II}^{(\xi_M\xi_M)}
    \end{bmatrix}.
\end{equation}
Here, each block $h_\mathrm{II}^{(\xi_i\xi_{i^\prime})}$ (consisting of a $3\times 3$ matrix) is obtained by performing a set of lattice sums. In particular, we have that the matrix elements of a specific block are given by
\begin{equation}
    \left[h_\mathrm{II}^{(\xi\xi^\prime)}\right]_{\beta\beta^\prime}=
    -\frac{3\pi\gamma_0}{k_0}\,e^{-i\kk(\bm{\delta}_\xi-\bm{\delta}_{\xi^\prime})}
    \sum_{\RR}G_{\beta\beta^{\prime}}(\bm{\delta}_\xi-\bm{\delta}_{\xi^\prime},\RR,k_0)\,e^{i\kk\RR}
\end{equation}
where, for the sake of clarity, we have made the notation change $\xi_i$ ($\xi_{i^\prime}$) $\rightarrow$ $\xi$ ($\xi^\prime$). These lattice sums are, in general, slowly convergent owing to the long-range character of the emergent photon-mediated interactions, but their numerical convergence can be accelerated by using the Ewald method~\cite{linton2010,campione2012}.

Finally, we diagonalize the effective Bloch Hamiltonian $H_\mathrm{eff}(\kk)$ by introducing a $\kk$-dependent similarity transformation $P(\kk)$. Then, we have that
\begin{equation}
    E(\kk) = P(\kk)^{-1}H_\mathrm{eff}(\kk)P(\kk),
\end{equation}
where $E(\kk)$ is, usually, a complex diagonal matrix from which one can identify the dispersive and dissipative components of the atomic array eigenmodes as:
\begin{align}
& \omega(\kk) = \mathrm{Re}\,E(\kk),
    \\[1.0ex]
& \gamma(\kk) = -2\,\mathrm{Im}\,E(\kk).
\end{align}
The diagonal elements of these matrices define the cooperative frequency shifts and collective decay rates of the Bloch modes composing the different bands of the system along the quasimomentum values $\kk$ that are chosen to belong to the first Brillouin zone. The effective Hamiltonian $\hat{H}_\mathrm{eff}$ can therefore be expressed as
\begin{equation}
    \hat{H}_\mathrm{eff} = \sum_{\nu,\kk}
    \left(\omega_{\nu\kk}-i\frac{\gamma_{\nu\kk}}{2}\right)
    \ket{\psi_{\nu\kk}}\bra{\psi_{\nu\kk}}
\end{equation}
where $\ket{\psi_{\nu\kk}}$ define a Bloch state. The latter can be expanded in the basis of the sublattice and polarization orbitals as follows:
\begin{equation}
\label{eq:Bloch_state_expansion}
    \ket{\psi_{\nu\kk}}=
    \sum_\xi\sum_\beta^{x,y,z}
    c_{\xi,\beta}\;\ket{\phi_{\xi\beta}}
\end{equation}
where the complex coefficients $c_{\xi,\beta}$ can be extracted from the matrix elements of the similarity transformation $P(\kk)$ that diagonalizes the effective Hamiltonian and we define $\ket{\phi_{\xi\beta}}=\hat{b}_{\kk\xi\beta}^\dagger\ket{\Psi_\text{vac}}$, with $\ket{\Psi_\text{vac}}$ denoting the overall vacuum state.

\begin{figure}[ht]
	\centering
	\includegraphics[width=16.5cm]{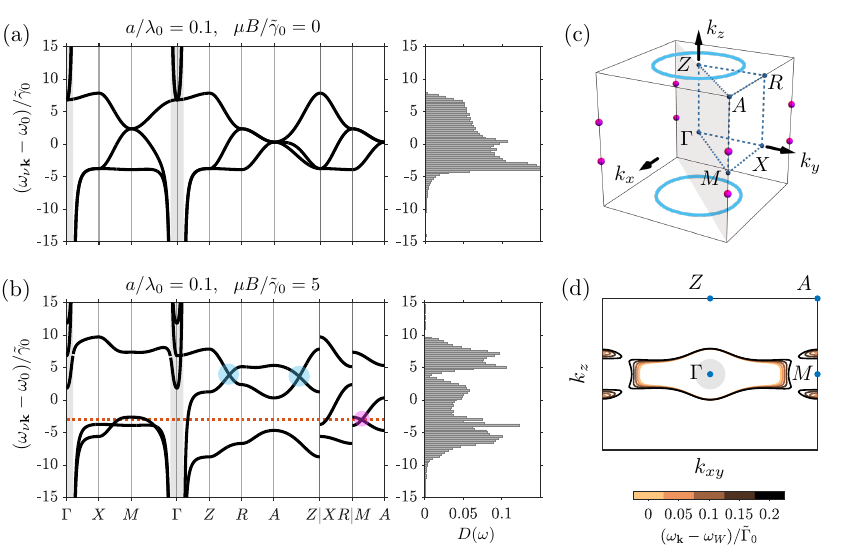}
	\caption{
       {\bf Emergence of Weyl excitations in simple cubic geometries.}
        (a)~Band structure of the atomic array in the absence of an external magnetic field. 
        (b)~Band structure of the atomic array in the presence of an external magnetic field giving rise to a Zeeman splitting value of $\mu B/\tilde{\Gamma}_0=5$.
        (c)~First Brillouin zone of the studied dipolar lattice. The cyan line denotes the localization of a nodal line in the CUB case. The magenta dot marks the position of the Weyl point highlighted with a magenta circle.
        (d,h)~Equifrequency contours calculated for several frequency values above the Weyl frequency along the $k_x-k_y=0$ plane.
	   \vspace{-0.3cm}}
	\label{fig:FigS1}
\end{figure}

\section{Energy dispersion of the simple cubic subwavelength atomic array~\label{sec:comparison}}

In this section, we provide a detailed survey of the polaritonic band structure associated with the simple cubic (CUB) configuration for a fixed interatomic distance of $a/\lambda_0=0.1$. The main results are summarized in Fig.~\ref{fig:FigS1}. 
In particular, the left panels of Figs.~\ref{fig:FigS1}(a) show the bands of the CUB lattice calculated along a high-symmetry path in the absence of any external magnetic field. The gray shadowed area around the $\Gamma$ point defines the light cone. 
The right panel of Fig.~\ref{fig:FigS1}(a) displays a discretized version of the associated density of states $D(\omega)$ which in the continuum limit yields
\begin{equation}
    D(\omega)=\sum_\nu\int_\mathrm{BZ}\frac{d^3\kk}{V_\mathrm{BZ}}\delta\left(\omega-\omega_{\nu\kk}\right).
\end{equation}

The fact that both inversion and time-reversal symmetry are preserved in the absence of an external magnetic field precludes the existence of a Weyl phase, but the situation changes drastically when the magnetic field is turned on. The latter breaks time-reversal symmetry, thereby enabling the emergence of Weyl excitations. Several point-like linear degeneracies are found, as shown in Figs.~\ref{fig:FigS1}(b). However, a more thorough inspection must be carried out to determine whether these correspond to Weyl points.

First, we examine the two linear crossings highlighted using cyan circles in Fig.~\ref{fig:FigS1}(b). Both of them occur at $\kk$-points belonging to the $k_z=\pm\pi/a$ plane, specifically in the $ZR$ and $AZ$ high-symmetry lines. Although they feature slightly different frequency values, they are essentially centered around $(\omega_{\nu\kk}-\omega_0)/\tilde{\Gamma}_0\approx 3.7$. A parabolic band with its minimum at the $\Gamma$ point is also observed within the same frequency window. It is worth noting that this feature prevents the studied crossings from being frequently isolated. However, such a condition can be achieved by engineering the $a/\lambda_0$ ratio and Zeeman splitting values.  Interestingly, the investigation of the dispersion relation over the $k_za=\pi$ plane reveals that these points belong to a nodal line, i.e., 1D ring-shaped degeneracy in the 3D Brillouin zone~\cite{burkov2011a}. The collection of points where this band degeneracy takes place is represented in Fig.~\ref{fig:FigS1}(c) using a solid cyan line. This exotic semimetallic phase has been intensively studied in condensed matter~\cite{bian2016,hu2016}, photonics~\cite{gao2018}, and phononics~\cite{deng2019}. Among other interesting phenomena, it can give rise to the so-called drumhead surface states in the form of flat bands~\cite{burkov2011a}. Here, we show that 3D atomic arrays under weak magnetic fields provide an ideal platform to test this kind of gapless topological phase. 

Next, we analyze the band crossing marked with a magenta dot in Fig.~\ref{fig:FigS1}(b). In particular, we show that the latter is unequivocally identified as a type I Weyl point. To prove that, one can study the equifrequency contours that are spectrally close to the Weyl frequency value $\omega_W$, represented with a red dotted line.
A careful inspection of such equifrequency contours demonstrates the presence of several close curves at the edges of the Brillouin zone which collapse to the Weyl points' positions indicated with magenta dots in Fig.~\ref{fig:FigS1}(c). This is illustrated in Fig.~\ref{fig:FigS1}(d) where we plot the equifrequency contours corresponding to $(\omega_{\nu\kk}-\omega_W)/\tilde{\Gamma}_0\in [0,0.2]$ over the $k_x-k_y=0$ plane, highlighted in gray in Fig.~\ref{fig:FigS1}(c).
Two Weyl points are found at the considered frequency value: one of them in the segment connecting the $M$ and the $A$ high-symmetry points and the other in the same vertical line but with an opposite sign of the $k_z$ coordinate. It is worth noting, however, that the emergent Weyl points are not frequency-isolated. This is clearly seen in Fig.~\ref{fig:FigS1}(d) where we find that in addition to the Weyl pockets, i.e., the closed curves that collapse to the Weyl points' positions, some extra equifrequency contours are present. The latter ones enclose the light cone, represented by a shadow gray circle, and survive as $\omega_{\nu\kk}$ approaches $\omega_W$ (see light brown line). Moreover, it is worth noting that the presence of the bands bending downwards at the edge of the light cone region, which can be interpreted as polaritonic states with a strong photonic component, precludes the occurrence of the frequency-isolated Weyl phase irrespective of the values chosen for $a/\lambda_0$ and $\mu B/\tilde{\gamma}_0$. 

\section{Movement of the Weyl points in reciprocal space~\label{sec:movement}}

In this section, we characterize the spectral properties of the system as a function of the interatomic distance $a$ and the Zeeman splitting $\mu B$ values. We show that a continuous swap of the studied parameters leads to smooth variations in the position and frequency of the Weyl points. This information can be used to determine the regions in the parameters' space wherein the band structure of the atomic array features a frequency-isolated pair of Weyl nodes.

\begin{figure}[p]
\centering
\includegraphics[width=16cm]{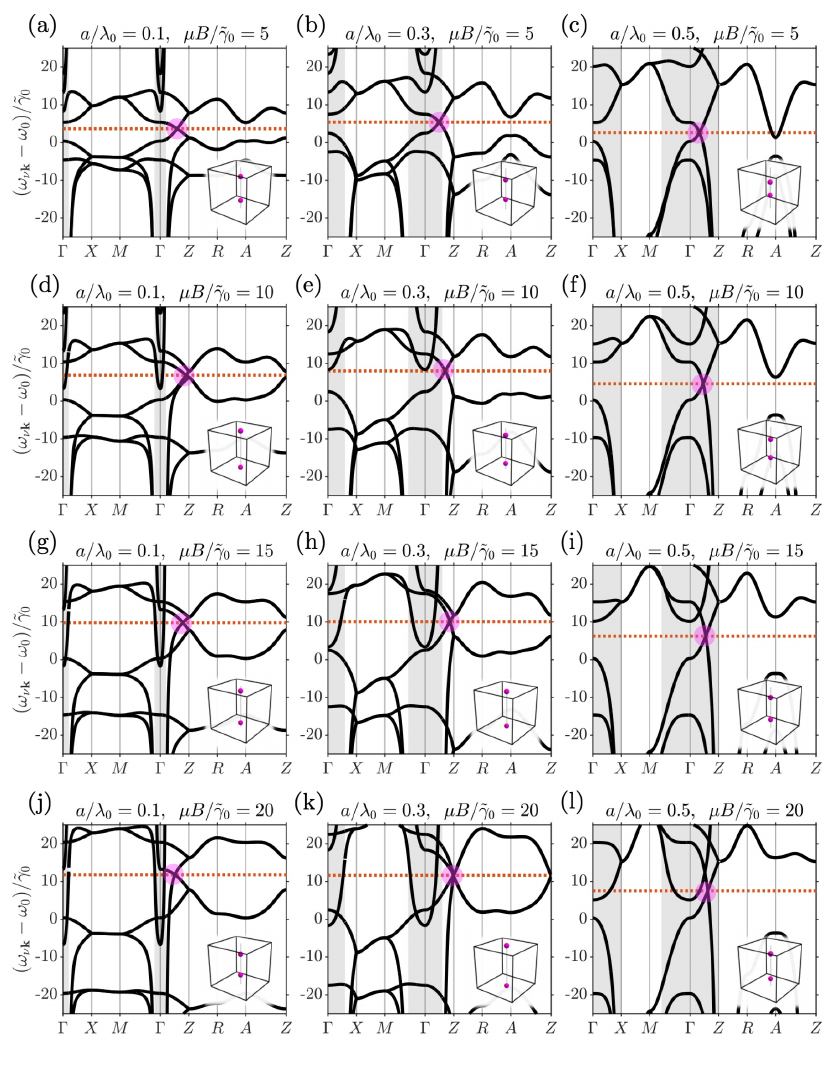}
\caption{
    {\bf Movement of the Weyl points in the body-centered cubic atomic array.}
    (a-l)~Band structure of the body-centered cubic array along a reduced high-symmetry path over the first Brillouin zone defined in Fig.~1(b) of the main text. Each panel corresponds to a given configuration with a fixed $a/\lambda_0$ ratio and Zeeman splitting $\mu B$ value. We highlight the linear crossing corresponding to the Weyl point with a positive $k_z$ coordinate using a magenta circle. The inset panels show the position of the Weyl points in reciprocal space. 
    }
\label{fig:FigS2}
\end{figure}

In Fig.~\ref{fig:FigS2} we depict the dispersion relation associated with the BCC lattice for several $a/\lambda_0$ and $\mu B/\tilde{\gamma}_0$ ratios. Each column corresponds to a given interatomic distance, whereas different rows correspond to distinct Zeeman splitting values.
For all the studied configurations, we find two Weyl singularities whose positions within the first Brillouin zone are highlighted using magenta dots in the inset panels. They appear in the line that defines the vertical axis, equidistant from the $\Gamma$ point, and with their $k_z$ coordinates featuring opposite signs. On the other hand, the Weyl frequency $\omega_W$ at which these degeneracies appear is marked with a red dotted line. 

Let us first examine the top row, which corresponds to a fixed Zeeman splitting value of $\mu B/\tilde{\gamma}_0=5$ and interatomic distances of $a/\lambda_0=0.1,0.3$ and $0.5$ (from left to right). The inset figures show that the Weyl points move toward the center of the Brillouin zone as larger values of the $a/\lambda_0$ ratio are considered. But more important is the fact that the light cone (shadow gray area) occupies a larger region of the first Brillouin zone as the interatomic separation is increased. Interestingly, the linear crossing observed in the $\Gamma Z$ symmetry line, which corresponds to the Weyl node with positive $k_z$ coordinate, reaches the boundary of the light cone for $a/\lambda_0=0.3$ and lies inside it for $a/\lambda_0=0.5$. 
A similar displacement of the Weyl points' positions is also observed for the second row (corresponding to the $\mu B/\tilde{\gamma}_0=10$ case), but not for the third and fourth rows (accounting for the $\mu B/\tilde{\gamma}_0=15$ and $\mu B/\tilde{\gamma}_0=20$ Zeeman splitting values, respectively). In the last two cases, the distance between the Weyl degeneracies and the $\Gamma$ point is smaller for $a/\lambda_0=0.1$ than for $a/\lambda_0=0.3$.
Yet, it is worth noting that, irrespective of the particular way the Weyl points move in reciprocal space, one can find some critical $a/\lambda_0$ ratio up to which the Weyl points fall inevitably into the light cone region. This allows us to divide the parameters' space into two regions depending on whether the Weyl points lie inside or outside the light cone. 

Now, we study how the position of the Weyl points changes as a function of the applied magnetic field for a fixed interatomic distance. If we focus on the left column, which illustrates the $a/\lambda_0=0.1$ case, we note that, as we start increasing the Zeeman splitting from the $\mu B/\tilde{\gamma}_0=5$ value, the Weyl points shift towards the top and bottom facets of the cubic Brillouin zone. At some particular value, which is close to $\mu B/\tilde{\gamma}_0\approx11$ for the $a/\lambda_0=0.1$ case, the Weyl nodes reach the $Z$ and $-Z$ points. If we continue increasing the magnetic field after that, the Weyl degeneracies move back toward the $\Gamma$ point. 
An intriguing scenario is the one in which the Weyl nodes are exactly at the $Z$ and $-Z$ points. Then, a four-fold degeneracy occurs at these high-symmetry locations, as seen, e.g., in Fig.~\ref{fig:FigS2}(k). The latter can be lifted by breaking the sublattice symmetry, which can be done by assuming two different atomic species with slightly distinct resonant transitions. In fact, a similar procedure has been proposed to construct 3D topological insulators with large Chern numbers in a purely photonic context~\cite{devescovi2021}.

The position of the Weyl points is not the only quantity that changes as we move around in the parameters' space. The frequency value at which the Weyl points appear is also modified. In particular, each column in Fig.~\ref{fig:FigS2} shows that, for all the studied interatomic distances, $\omega_W$ increases monotonically as a larger value of the Zeeman splitting is assumed. Also, the bands that are not directly related to the linear crossing giving rise to the Weyl degeneracies evolve as different combinations of the $a/\lambda_0$ ratio and $\mu B$ value are considered. It can happen then that some of the bands intercept the Weyl frequency, thereby precluding the occurrence of the ideal Weyl phase.

\section{Surface modes confinement~\label{sec:surface}}

\begin{figure}[h]
	\centering
	\includegraphics[width=16.5cm]{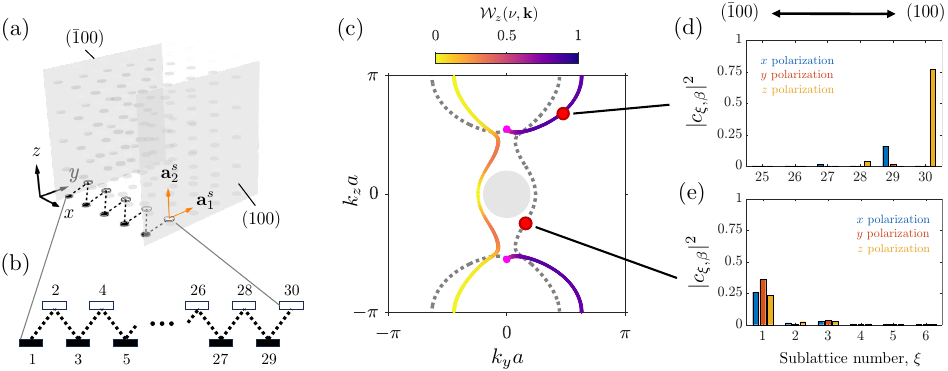}
	\caption{
       {\bf Confinement of the subradiant Fermi arc surface states.}
       (a)~Squematics of the employed slab model. 
       (b)~ Labeling of the different sublattices forming the unit cell in the slab. 
       (c)~Equifrequency contours corresponding to the Weyl frequency for the subradiant slab bands provided that the system is prepared in the configuration in which $a/\lambda_0=0.1$ and $\mu B/\tilde{\gamma}_0=5$ [see Fig~3(d) of the main text].
       (d,e)~Localization of the Bloch states forming the Fermi arcs.
	   \vspace{-0.3cm}}
	\label{fig:FigS3}
\end{figure}

To determine the localization of the Bloch modes composing the Fermi arcs we leverage the expansion of the Bloch state $\ket{\psi_{\nu\kk}}$ in terms of the sublattice and polarization orbitals introduced in Eq.~\eqref{eq:Bloch_state_expansion}. It is worth noting that any slab model can be thought of as a two-dimensional lattice with an extended unit cell spanning the transversal direction, i.e., the one along which the system is assumed to be finite. The appropriate labeling of the non-equivalent sites that form the unit cell [see Fig~\ref{fig:FigS3}(b)] allows us to identify the orbitals that are localized in the $\bar{1}00$ and $100$ facets. Since we consider a unit cell with $M=30$ non-equivalent sites, we have that orbitals labeled as $\xi\approx1$ are located close to the $\bar{1}00$ surface whereas those labeled as $\xi\approx30$ are close to the $100$ facet.

Following Eq.~\eqref{eq:Bloch_state_expansion}, we note that any given Bloch state, with well-defined band number $\nu$ and quasimomentum $\kk$, can be expanded as follows:
\begin{equation}
    \ket{\psi_{\nu\kk}}=
    \sum_{\xi=1}^M\left(
    c_{\xi,x}\ket{\phi_{\xi x}}+
    c_{\xi,y}\ket{\phi_{\xi y}}+
    c_{\xi,z}\ket{\phi_{\xi z}}
    \right),
\end{equation}
where we have explicitly separated the three polarization components. 
We can calculate the projection of $\ket{\psi_{\nu\kk}}$ over the sublattice $\xi$ and polarization $\beta$ as $|c_{\xi\beta}|^2$. In Figs~\ref{fig:FigS3}(d) and (e), we plot this quantity for two particular Bloch states belonging to the Fermi arcs as a function of the sublattice number. By doing so, we find that the state plotted in Fig~\ref{fig:FigS3}(d) features a strong localization over the $100$ facet, whereas the one represented in Fig~\ref{fig:FigS3}(e) is localized in the opposite surface. The distinct color bars denote the weight of each polarization for a given sublattice. By summing over sublattices, we obtain
\begin{equation}
    \mathcal{W}_{\beta}(\nu,\kk)=\sum_{\xi=1}^M|\braket{\phi_{\xi \beta}}{\psi_{\nu\kk}}|^2=\sum_{\xi=1}^M|c_{\xi\beta}|^2,
\end{equation}
that defines the total weight of the polarization $\beta$ in the the Bloch state $\ket{\psi_{\nu\kk}}$. 

\begin{thebibliography}{70}
\bibitem [{foo()}]{footnote1}%
  \BibitemOpen
  \href@noop {} {\bibinfo {title} {{We also note that inversion symmetry is
  preserved which guarantees that Weyl points with opposite quasimomentum
  values have opposite chirality~\cite{delplace2012a}.}}}\BibitemShut {Stop}%
\end{thebibliography}
\end{document}